\documentclass[twocolumn]{IEEEtran}
\usepackage{color}
\usepackage{graphicx}
\usepackage{amsmath}
\usepackage{amsthm}
\usepackage{cite}
\usepackage[caption=false,font=normalsize,labelfon
t=sf,textfont=sf]{subfig}
\usepackage{subfig}
\usepackage{dsfont}
\usepackage{stfloats}
\usepackage[letterpaper, margin=0.65in, top=0.75in]{geometry}
\usepackage{epsfig,epstopdf,amssymb,enumerate} 

\begin{document}

\title{ A Novel 2D Non-Stationary Wideband Massive MIMO Channel Model}
\author{Carlos F. Lopez$^1$, Cheng-Xiang Wang$^1$, and Rui Feng$^2$\\

$^1$\small{Institute of Sensors, Signals and Systems, Heriot-Watt University, Edinburgh EH14 4AS, UK.} \\
$^2$\small{Shandong Provincial Key Lab of Wireless Communication Technologies, Shandong University, Jinan, Shandong, 250100, China.} \\
Email: \{c.f.lopez, cheng-xiang.wang\}@hw.ac.uk, fengxiurui604@163.com }

\maketitle

\begin{abstract}
In this paper, a novel two-dimensional (2D) non-stationary wideband geometry-based stochastic model (GBSM) for massive multiple-input multiple-output (MIMO) communication systems is proposed. Key characteristics of massive MIMO channels such as near field effects and cluster evolution along the array are addressed in this model. Near field effects are modeled by a second-order approximation to spherical wavefronts, i.e., parabolic wavefronts, leading to linear drifts of the angles of multipath components (MPCs) and non-stationarity along the array. Cluster evolution along the array involving cluster (dis)appearance and smooth average power variations is considered. Cluster (dis)appearance is modeled by a two-state Markov process and smooth average power variations are modeled by a spatial lognormal process. Statistical properties of the channel model such as time autocorrelation function (ACF), spatial cross-correlation function (CCF), and cluster average power and Rician factor variations over the array are derived. Finally, simulation results are presented and analyzed, demonstrating that parabolic wavefronts and cluster soft evolution are good candidates to model important massive MIMO channel characteristics.

\vspace{0.1cm}
\end{abstract}

\IEEEpeerreviewmaketitle

\section{Introduction}
Massive MIMO communication technologies have been proposed as a key enabler to address important challenges of the fifth generation (5G) wireless communication systems. The introduction of many antennas has demonstrated improving both efficiency and reliability of wireless communication systems. For example, the capacity of steering sharper beams and, consequently, using a higher array gain enable to reduce interference and energy consumption. Also, more degrees of freedom permits increasing both spectral efficiency and reliability through spatial multiplexing and diversity schemes, respectively \cite{Larsson2013, Persson2012}.

Increasing the number of antennas may result in arrays covering large distances, often beyond the stationary interval of the channel. This leads to new channel characteristics that need to be modeled for the accurate performance evaluation of wireless communication systems. Recent measurements have demonstrated the so called {\it near field effects} due to the presence of scatterers and users in the near field region of the array, and environment variations as perceived by different antenna elements\cite{Payami2012,Gao2012,Gao2013}. 

Recently, significant effort has been made to model these effects. The authors in \cite{Wu2014} and \cite{Wu2015a} used spherical wavefronts to account for near field effects and a birth-death process for environment variations over the array. In these models, spherical wavefronts and cluster (dis)appearance result in non-linear angular variations of the MPCs and sudden variations of the total received power along the array, respectively. 
In \cite{Gao2013}, a different approach was used to model environment variations along the array by extending the concept of visibility region (VR) introduced in COST 2100 channel model \cite{Liu2012} to the base station (BS) large array. In this approach, BS-VRs define subsets of consecutive antennas where clusters are visible. Contrary to the approach in \cite{Wu2014} and \cite{Wu2015a}, the authors in \cite{Gao2013} proposed a {\it visibility gain}, a  method that enables to model linear power variations over the array. However, these BS-VRs are defined by segments and do not consider the possibility of segmented VRs on the array where clusters reappear.

Measurements \cite{Payami2012, Gao2012, Gao2013,Li2015} have shown that near field effects often lead to linear drifts of the angles of MPCs in many practical situations. Also, a more sophisticated evolution of clusters along the array compared to what was proposed in previous channel models \cite{Wu2014,Wu2015a, Gao2013} can be observed. As indicated by the same authors, a cluster shadow fading process to model power variations over the large array may be necessary. Finally, it has been pointed out in \cite{Wang2016b} that current massive MIMO channel models involve an undesirable high computational complexity. 

In this paper, a 2D non-stationary wideband GBSM inspired by WINNER models \cite{WINNERII-models} capable of capturing key massive MIMO channel  characteristic is proposed. In this model, a second-order approximation to spherical wavefront, i.e., parabolic wavefront, is introduced. This approximation results in linear angular drifts over the array, reducing theoretical and computational complexity compared to spherical wavefronts \cite{Swind88}. Moreover, a more flexible and accurate cluster evolution approach is proposed by introducing independent Markov processes per cluster and modeling cluster average power variations along the array through spatial lognormal processes. 

The rest of the paper is organized as follows. Section \ref{sec:ChannelModel} describes the novel massive MIMO channel model proposed, including parabolic wavefronts and cluster evolution along the array. In Section \ref{sec:StatisticalProperties}, statistical properties of the model such as the time ACF, spatial CCF, cluster average power, etc. are derived. In Section \ref{sec:Simulations}, simulation results are presented and analyzed. Finally, conclusions and limitations of the model are drawn in Section \ref{sec:conclusions}. 

\section{A novel Massive MIMO channel model}
\label{sec:ChannelModel}
\begin{figure}[htp]
\centerline{
\includegraphics[width=0.5\textwidth]{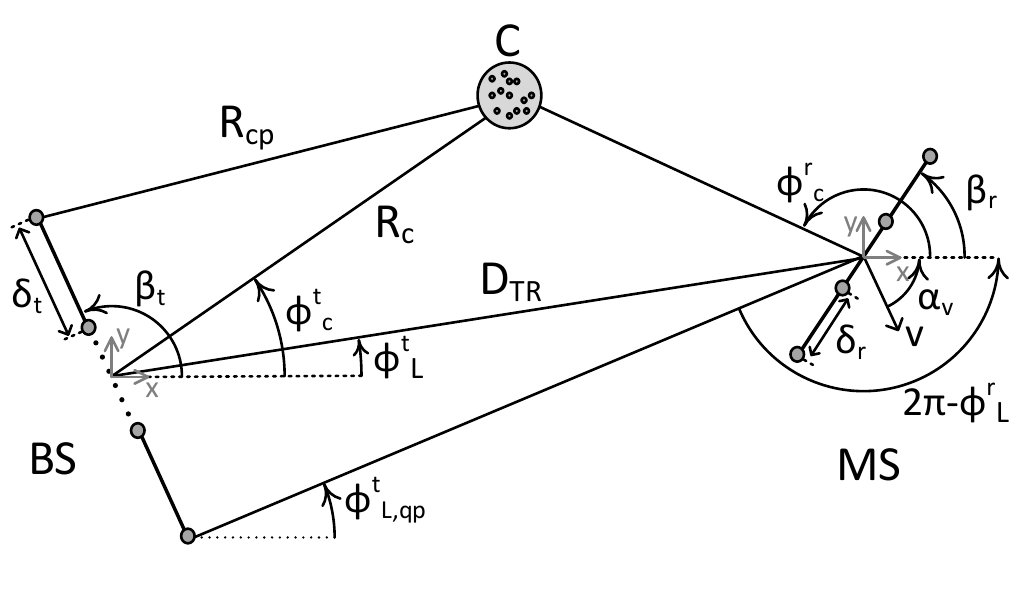} }

\caption{ The massive MIMO channel model.}
\label{fig:channel}
\end{figure}

In this channel model, only the base station (BS) array is considered large compared to the wavelength and the stationary interval, and only the mobile station (MS) is in motion. From now on, the BS is considered as the transmitter and the MS as the receiver. 
In Fig. \ref{fig:channel}, a schematic model of the channel is presented. The BS uniform linear array (ULA) is formed by $M_t$ antenna elements equally spaced at a distance $\delta_t$ and its axis is tilted an angle $\beta_t$ with respect to (w.r.t.) the x-axis. The MS is a ULA formed by $M_r$ antenna elements equally spaced at a distance $\delta_r$ and tilted an angle $\beta_r$ w.r.t. the x-axis. The reference point for both arrays is at the center of each and the $c^{th}$ cluster is denoted by $\text{C}$ in the figure. Since the MS array is not large, the far field assumption is used for this side of the channel. Other model parameters such as angle of arrival (AoA), angle of departure (AoD), etc. are defined in Table \ref{tab:parameters}.  
\vspace{-0.3cm}

\begin{table*}[ht]
\caption{Definition of parameters of the massive MIMO channel model in Fig. \ref{fig:channel}.}
\vspace{-0.2cm}
\center
\label{tab:parameters}
\begin{tabular}{|c|l|l|}
\hline
{\bf Parameter}  & {\bf Definition}          \\ \hline
$R_c$                 &  Distance between the $c^{th}$ cluster and the center of the BS array.          \\ \hline
$R_{c,p}$            &  Distance between the $c^{th}$ cluster and the $p^{th}$ BS antenna element.   \\ \hline
$D_{TR}$           & Distance between the centers of the BS and MS arrays.\\ \hline
$\delta_t,\delta_r$           & Distance between consecutive elements in the BS and MS arrays, respectively. \\ \hline
$\beta_t,\beta_r$     & Tilt angles of the BS and MS arrays w.r.t. x-axis, respectively. \\ \hline
$\phi^{t}_L,\phi^{r}_L$ & AoD/AoA of the line of sight (LOS) rays measured from the center of the BS/MS arrays.\\ \hline
$\phi_{L,qp}^{t},\phi_{L,qp}^{r}$ & AoD/AoA of the LOS rays from BS antenna $p$ to MS antenna $q$.\\ \hline
$\phi_{c}^t,\phi_{c}^r$ & Azimuth coordinate of the $c^{th}$ cluster measured from the center of the BS/MS arrays.\\ \hline
$\phi_{m_c}^t,\phi_{m_c}^r$ & AoD/AoA of the $m^{th}$ MPC bounced by the $c^{th}$ cluster measured from the centers of the BS/MS arrays.\\ \hline

$\alpha_v$               & Angle of MS velocity w.r.t. the x-axis.\\ \hline
$v$ 	                      & Speed of MS.\\ \hline
\end{tabular}
\end{table*}

The wideband MIMO channel is represented by a matrix ${\ {\bf H}}(t,\tau)=[h_{qp}(t,\tau)]_{M_r \times M_t}$ where $p=1,2, \dots, M_t$ and $q=1,2, \dots, M_r$. The channel impulse response (CIR) $h_{qp}(t,\tau)$  is calculated as the superposition of the line-of-sight (LOS) and non-LOS (NLOS) components as follows:
\begin{equation}
h_{qp}(t,\tau) =h_{L,qp}(t)\delta(\tau - \tau_1) + \sum_{c=1}^{C} { h_{c,qp}(t) \delta(\tau - \tau_c)}
\label{eq:CIR}
\end{equation}
where the excess delays $\tau_c$ are associated to the MPCs of the $c^{th}$ cluster. The LOS and NLOS components of the channel impulse response in (\ref{eq:CIR}) are defined as
\begin{eqnarray}
\label{eq:CIR_LOS}
h_{L,qp}(t) &=&\sqrt{P_{L,p}} e^{j(\Delta\Phi_{L,qp}+2\pi f_{L,p} t) } \\
h_{c,qp}(t) &=&\sqrt{\frac{P_{c,p}}{M_c}} \sum_{m_c=1}^{M_c}{\!\!\!e^{j\theta_{m_c}}e^{j(\Delta\Phi_{m_c,qp}+2\pi f_{m_c,qp} t) }} 
\label{eq:CIR_c}
\end{eqnarray}
where $M_c$ is the number of delay-irresolvable subpaths or rays per cluster and $\theta_{m_c}$ are independent and identically distributed (i.i.d.) random variables uniformly distributed over the interval $[0,2\pi]$ defining the phases of the $m^{th}$ MPC of the $c^{th}$ cluster at the center of the BS array. The parameters $\Delta\Phi_{L,qp}$ and $\Delta\Phi_{m_c,qp}$ are the phase changes experienced by the LOS component and the $m^{th}$ MPC scattered by the $c^{th}$ cluster from the transmitter antenna $p$ to the receiver antenna $q$, respectively. The parameters $P_{L,p}$ and $P_{c,p}$ denote the powers of the LOS component and the cluster average power at the BS antenna element $p$, respectively. The parameter $f_{m_c}$ indicates the Doppler shifts of the MPCs for every cluster. The parameters $\Delta\Phi_{L,qp}$, $\Delta\Phi_{m_c,qp}$, $P_{L,p}$, $P_{c,p}$, $f_{L,p}$ and $f_{m_c}$ will be described in Sections \ref{sec:ParabolicWavefronts} and \ref{sec:PowerVariationArray}.

\subsection{Parabolic wavefronts: second-order approximation}
\label{sec:ParabolicWavefronts}
One cause of non-stationarity of the channel statistical properties along the array are the near field effects. Since clusters or MSs might be located within the Fresnel region, spherical wavefronts haven been considered in the state-of-the-art massive MIMO channel models\cite{Wu2014,Wu2015a,Gao2013}. In our model, for the reasons previously indicated in the introduction, parabolic wavefronts instead of spherical wavefronts are used. 

With reference to Fig. \ref{fig:channel}, it is assumed that the positions of the clusters and MS are defined by their polar coordinates $(R_c, \phi_c^t)$ and $(D_{TR},\phi_L^t)$, respectively. The distance from the $c^{th}$ cluster to the $p^{th}$ antenna element of the BS array is given by the law of cosines as follows:
\begin{equation}
\begin{split}
\label{eq:Cdistance}
R_{c,p}^2 =&R_c^2 + (M_t-2p+1)^2(\delta_t/2)^2 \\
&-(M_t-2p+1)\delta_t R_c\cos(\phi_c^t-\beta_t).
\end{split}
\end{equation}
The second-order (parabolic) approximation to (\ref{eq:Cdistance}) is 
\begin{equation}
\begin{split}
R_{c,p}\approx R_c - (M_t-2p+1) (\delta_t/2)\cos(\phi_c^t-\beta_t) \\
+  [(M_t-2p+1)(\delta_t/2)]^2 \frac{\sin^2(\phi_c^t-\beta_t)}{2 R_c}.
\end{split}
\label{eq:fresnel_a}
\end{equation}

Using the previous approximation (\ref{eq:fresnel_a}), the phase difference $\Delta\Phi_{m_c,qp}$ in (\ref{eq:CIR_c}) can be computed through the difference in distance between the link $p-C-q$ and a reference link connecting the centers of both arrays via the $c^{th}$ cluster, as it is usually computed \cite{WINNERII-models,Patzold2012}. This can be expressed as the sum of two terms representing the planar wavefront $\Delta\Phi_{m_c,qp}^{PL}$ and parabolic wavefront $\Delta\Phi_{m_c,qp}^{PB}$ defined as
\begin{align}
\label{eq:PLphaseDiff_c}
\nonumber
\Delta\Phi_{m_c,qp}^{PL} =& \: \kappa[(M_t\!-\!2p\!+\!1) (\delta_t/2) \cos(\phi_{m_c}^t-\beta_t) \\
				    &\:\:\:\:\:\:\:\:\:+(M_r\!-\!2q\!+\!1) (\delta_r/2) \cos(\phi_{m_c}^r-\beta_r)]\\ 
\Delta\Phi_{m_c,qp}^{PB} =& -\kappa [(M_t\!-\!2p\!+\!1)(\delta_t/2)]^2 \frac{\sin^2(\phi_{m_c}^t-\beta_t)}{2 R_c}
\label{eq:PBphaseDiff_c}
\end{align}
where $\kappa = 2\pi/\lambda$ and $\lambda$ is the carrier wavelength. On the one hand, note that (\ref{eq:PLphaseDiff_c}) represents the plane-wave approximation used in conventional non-massive MIMO channel models. On the other hand, (\ref{eq:PBphaseDiff_c}) introduces the second-order approximation to the spherical wavefront, i.e, the parabolic wavefront, that vanishes for large distances $R_c$, i.e., when the far-field assumption is applied. The phase difference $\Delta\Phi_{L,qp}$ in (\ref{eq:CIR_c}) can be computed analogously by substituting $\phi_{m_c}^{r/t}$ by $\phi_{c}^{r/t}$ in (\ref{eq:PLphaseDiff_c}) and (\ref{eq:PBphaseDiff_c}) and $R_c$ by $D_{TR}$ in (\ref{eq:PBphaseDiff_c}). For specular scattering , not considered in this model, the distance to the source through its images instead of the distance to the last scatterer must be considered for (\ref{eq:PBphaseDiff_c}) to be valid.

Due to the parabolic wavefront, (\ref{eq:CIR_c}) does not represent a sum-of-cisoids process in the BS antenna index $p$, but a sum-of-chirp or linear-frequency-modulated signals in that domain. Consequently, it can be demonstrated that the angles of the MPCs linearly drift over the array \cite{Swind88, Patzold16}. 

The Doppler shifts experienced by the LOS and MPCs from the $c^{th}$ cluster are computed as
\begin{eqnarray}
\label{eq:fdopplerLOS}
f_{L,qp} &=& f_{max} \cos(\phi_{L,qp}^r-\alpha_v) \\
f_{m_c} &=& f_{max} \cos(\phi_{m_c}^r-\alpha_v)
\label{eq:fdopplerNLOS}
\end{eqnarray}
where $f_{max}=\lVert v \lVert /\lambda$ is the maximum Doppler shift. Since linear drift of the angle has been considered along the BS large array, the angle of the LOS component $\phi_{L,qp}^r$ in (\ref{eq:fdopplerLOS}) can be approximated as
\begin{equation}
\phi_{L,qp}^r = \pi +\phi_{L}^t + \sin(\phi_{L}^t-\beta_t) \frac{(M_t-2p+1)\delta_t}{2D_{TR}}.
\end{equation}
This approximation leads to an array-variant Doppler spectrum whose rate of change is constant for every MS and depends on the angle and distance to the MS considered. For distant MSs, $D_{TR}  \gg (M_t-2p+1)\delta_t$ and the Doppler spectrum becomes array-invariant, as it is usually considered in non-massive MIMO models. Note that the angles $\phi_{L,qp}^r$ and $\phi_{m_c}^r$ in (\ref{eq:fdopplerLOS}) and (\ref{eq:fdopplerNLOS}) do not depend on the index $q$ because the MS array is not large. Also, the index $p$ does not affect $\phi_{m_c}^r$ in (\ref{eq:fdopplerNLOS}) because all the rays bounced by the $c^{th}$ cluster experience the same Doppler shift regardless of the BS antenna element. As previously referred, these dependence relationships are correct for scattered multipath components, but specular reflections should be treated separately as the distance to the source through its images must be considered. 

\subsection{Smooth evolution of clusters along the array}
\label{sec:PowerVariationArray}
Variations of the cluster average power are introduced via two processes accounting for two related phenomena: cluster (dis)appearance and smooth cluster average power variations. The first phenomenon is modeled by a two-state Markov process in a similar way as it has been done in \cite{Wu2014} and \cite{Wu2015a}. The second effect is modeled by a spatial lognormal process, which is coherent with the underlying physical phenomena where both LOS and MPCs scattered by clusters can be partially occluded or shadowed. 
Moreover, as it will be shown below, it is a natural way of extending WINNER models through the randomization parameter of cluster average power \cite{WINNERII-models}.  

In WINNER models, the cluster average power $P_c$ is modeled as a function of the delay and the environment as
\begin{equation}
P_c  = e^{\left (-\tau_c \frac{r_\tau -1 }{r_\tau\sigma_\tau}\right)}\cdot 10^{\frac{-\nu_c}{10}}
\label{eq:ClusterPowerWINNNER}
\end{equation}
where $\tau_i$ is the delay of the signal scattered by the $i^{th}$ cluster, $\sigma_\tau$ is the delay spread of the channel and $r_\tau$ is a scenario dependent parameter representing the ratio of the standard deviation of the path delays and the root mean square (RMS) delay spread. The parameter $\nu_c$ is an i.i.d. Gaussian random variable with zero mean and standard deviation $\sigma_\nu=3$ introduced to model the shadowing randomization effect on each cluster for every simulation drop or segment \cite{WINNERII-models}.  The cluster average power is normalized before is introduced in (\ref{eq:CIR_c}) so that the sum of the power from all clusters equals unity. 

In the proposed massive MIMO model, in order to model mean power variations over the array, the factor $10^{\frac{-\nu_c}{10}}$ in (\ref{eq:ClusterPowerWINNNER}) is dropped and the following modification is proposed:
\begin{equation}
P_{c,p}  = P_c\cdot \xi_{c,p}^2 \cdot \Pi_{c,p}^2
\label{eq:ClusterPower}
\end{equation}
where the effects of the term $10^{\frac{-\nu_c}{10}}$ in (\ref{eq:ClusterPowerWINNNER}) have been replaced and extended by $(\xi_{c,p} \cdot\Pi_{c,p})^2$. The parameters $\xi_{c,p}$ and $\Pi_{c,p}$ are a discrete-spatial lognormal process used to model smooth cluster average power variations along the array, and a two-state discrete Markov chain to characterize cluster (dis)appearance along the array, respectively. Note that both processes depend on cluster and antenna indices. Also, smooth power variations of the LOS component and transitions LOS-NLOS along the large array are modeled analogously by the processes $\xi_{L,p}$, and $\Pi_{L,p}$, respectively. Thus, the power of the LOS component in (\ref{eq:CIR_LOS}) is $P_{L,p} =  (\xi_{L,p}\cdot \Pi_{L,p})^2$. 

Applying the concept of continuous spatial shadowing processes described in \cite{Patzold2012} to the BS array, the discrete spatial process $\xi_{c,p}$ can be obtained by sampling a continuous-spatial lognormal process $\xi_c(x)$ at the positions of every antenna element of the BS array as
\begin{equation}
\xi_{c,p} = 10^{(\sigma_{c} \nu_{c,p} +m_{c})/20}
\label{eq:DiscreteLogNormal}
\end{equation}
where $\nu_{c,p}= \nu_c(  (M_t-2p+1) \delta_t/2 )$ is a real-valued zero-mean spatial Gaussian wide-sense stationary (WSS) process with unit variance sampled as described above. The parameters $\sigma_c$ and $m_c$ are called the {\it shadow standard deviation} and the {\it area mean}, respectively, and they control the amplitude and standard deviation of the lognormal process. As indicated by the index $c$ in (\ref{eq:DiscreteLogNormal}), these parameters can be different for every cluster and dependent on the geometric characteristics of the environment. The parameter $m_{c}$ is a function of the distance to the cluster, frequency, and other parameters depending on the path-loss model applied. Regarding the standard deviation $\sigma_{c}$, it has been found in measurements that this parameter is usually in the range from 5 to 12 dB at 900 MHz, and 8 dB is a typical value for macrocellular applications \cite{Patzold2012}.

The spatial ACF of the Gaussian process  $\nu_{c,p}$ models the rate of change of the cluster average power over the array. If the Gaussian correlation model is used, this ACF is defined as
\begin{equation}
r_{\nu_{c,pp'}}(\delta_t)={\mathrm E}[ \nu_c(x_p) \nu_{c}^*(x_{p'})] = e^{-(\delta_t(p-p')/D_c)^2}
\label{eq:GaussCorr}
\end{equation}
with $x_i = (M_t-2i+1) \delta_t/2$ for $i=\{p,p'\}$. The parameter $D_c$ is called the {\it decorrelation distance}, defined as the distance from the origin where the ACF in (\ref{eq:GaussCorr}) becomes $e^{-1}$. Since the ACF of the process $\nu_{c,p}$ only depends on relative distances, it is WSS. The ACF of the process $\xi_{c,p}$ in (\ref{eq:DiscreteLogNormal})  $r_{\xi_{L/c,pp'}}(\delta_t)={\mathrm E}[\xi_{L/c,p}\xi_{L/c,p'}]$ is given by \cite{Patzold2012}
\begin{align}
\!\!\!r_{\xi_{L/c,pp'}}(\delta_t)=e^{2m_{L/c}+ \sigma_{L/c}^2[1+r_{\nu_{c,pp'}}(\delta_t)]}.
\end{align}
The notation of the subscripts $L/c$ indicates either the LOS component or the cluster component of the previous processes and correlation functions, respectively.

A two-state Markov chain $\Pi_c(x)$ models cluster appearance and disappearance over the array. Since every cluster is visible for a certain region along the array, $\Pi_c(x)$ takes values 1 and 0 depending on the cluster visibility over such dimension. The visibility and invisibility regions of a cluster are considered exponential i.i.d. random variables with intensities $\lambda_v$ and $\lambda_i$, respectively. The probability of transition between regions is defined by the transition matrix as \cite{Papoulis02}
\begin{eqnarray}
T_c(x) = \frac{1}{\lambda_T^c}  \left ( \! \begin{array}{cc}
\lambda_i^c + \lambda_v^c e^{-\lambda_T^cx}   &  \lambda_v^c -  \lambda_v^c e^{-\lambda_T^cx} \\
\lambda_i^c - \lambda_i^c e^{-\lambda_T^cx} &   \lambda_v^c +  \lambda_i^c e^{-\lambda_T^cx} \\
\end{array} \! \right )
\label{eq:MTP}
\end{eqnarray}

where $\lambda_T^c=\lambda_v^c+\lambda_i^c$. The probabilities that a cluster is visible or invisible at any position $x$ along the array are $p_v = \lambda_v^c/\lambda_T^c$, $p_i=\lambda_i^c/\lambda_T^c$, respectively. The Markov chain in (\ref{eq:ClusterPower}) is a sampled version of the continuous process $\Pi_{c}(x)$ at the positions of the BS antenna elements $\Pi_{c,p} = \Pi_c( (M_t-2p+1) \delta_t/2 )$. Note that the transition rates $\lambda_v^c$ and $\lambda_i^c$ might be different for every cluster. 
The ACF of the Markov two-state process $r_{\Pi_{L/c,pp'}}(\delta_t) = {\mathrm E}[\Pi_{L/c,p}\Pi_{L/c,p'}]$ can be expressed as \cite{Papoulis02}
\begin{align}
\!\!\!r_{\Pi_{L/c,pp'}}(\delta_t)=\frac{\lambda_v^{L/c}} {\lambda_v^{L/c}+\lambda_i^{L/c}} e^{-(\lambda_v^{L/c}+\lambda_i^{L/c})\delta_t (p-p')}.
\end{align}
Again, the ACF of the process $\Pi_{c,p}$ only depends on relative distances, so it is WSS as well.

\section{Statistical properties of the channel model}
\label{sec:StatisticalProperties}
Since it is assumed that the LOS and MPCs from different clusters are uncorrelated, the ACF of the channel $r_{qp}(t,\Delta t) = {\mathrm E}[h_{qp}(t) h_{qp}^*(t+\Delta t)]$ can be separated as 
\begin{equation}
r_{qp}(t,\Delta t,\tau)  = r_{L,qp}(\Delta t) + \sum_{c=1}^{C} {r_{c,qp}(t,\Delta t)\delta(\tau-\tau_c) }.
\label{eq:acf}
\end{equation}

The ACF of the LOS component $r_{L,qp}(\Delta t)$ and ACF per cluster component $r_{c,qp}(\Delta t)$ are given by 
\begin{eqnarray}
\label{eq:acf_LOS}
\hspace*{-6mm}r_{L,qp}(\Delta t) \!\!\!\!&=&\!\!\!\!{\mathrm E}[\xi_{L,p}^2] {\mathrm E}[\Pi_{L,p}^2] e^{-j2\pi f_{L,p} \Delta t} \\
\hspace*{-6mm}r_{c,qp}(\Delta t) \!\!\!&=&\!\!\!\! {\mathrm E}[\xi_{c,p}^2] {\mathrm E}[\Pi_{c,p}^2]\frac{P_{c}}{M_c} \sum_{m=1}^{M_c}{{\mathrm E}[ e^{-j2\pi f_{m_c,p}\Delta t}]} 
\label{eq:acf_c}
\end{eqnarray}
where it has been assumed that the processes $\xi_{c/L}$ and $\Pi_{c/L}$  are independent. Since the ACFs in (\ref{eq:acf_LOS}) and  (\ref{eq:acf_c}) are time invariant and only depend on time difference $\Delta t$, the model is WSS in time domain.  Also, note that (\ref{eq:acf_LOS}) and (\ref{eq:acf_c}) do not reflect the power variations along the array because the contribution of the large-scale processes $\xi_{c,p}, \Pi_{c,p}$ is averaged as well. However, the total power received by the BS antenna element $p$ can be computed as the sum of the LOS and cluster average power as defined in (\ref{eq:ClusterPower}) as
\begin{equation}
P_p = \xi_{L,p}^2 \cdot \Pi_{L,p}^2 + \sum_{c=1}^{C}{\xi_{c,p}^2 \cdot \Pi_{c,p}^2}
\label{eq:PowerVariations}
\end{equation}
showing that the mean power of the CIR is non constant along the array. Consequently, the Rice K factor that indicates the proportion of LOS to NLOS power is array variant as well and can be expressed as
\begin{equation}
K_p = \frac{\xi_{L,p}^2 \cdot \Pi_{L,p}^2} {\sum_{c=1}^{C}{\xi_{c,p}^2 \cdot \Pi_{c,p}^2}}.
\label{eq:Kfactor}
\end{equation}
Using the previous assumptions on the correlation between LOS and cluster MPCs, the spatial CCF $ \rho_{qp,q'p'}(\delta_t,\delta_r, \tau)= {\mathrm E}[h_{qp}(t) h_{q'p'}^*(t)]$ is given by
\begin{align}
\nonumber
\rho_{qp,q'p'}(\delta_t,\delta_r, \tau)  = & \:\rho_{L,qp,q'p'}(\delta_t,\delta_r) \\ 
+&\sum_{c=1}^{C} {\rho_{c,qp,q'p'}(\delta_t,\delta_r)\delta(\tau-\tau_c) }.
\label{eq:CCF}
\end{align}
The LOS and per cluster components of the spatial CCF are 
\begin{eqnarray}
\label{eq:CCF_LOS}
\!\!\!\!\!\!\!\!\!\!\!\rho_{L,qp,q'p'}\!\!\!\!&=& \!\!\!\!\! r_{\xi_{L,pp'}} r_{\Pi_{L,pp'}}e^{-j [\varphi_{L,qp,q'p'} + 2\pi t(f_{L,p}-f_{L,p'})]}  \\
\!\!\!\!\!\!\!\!\!\!\!\rho_{c,qp,q'p'} \!\!\!\!&=&\!\!\!\!\! r_{\xi_{c,pp'}} r_{\Pi_{c,pp'}} \frac{P_c}{M_c} \sum_{m=1}^{M_c}
{\mathrm E}[e^{-j \varphi_{c,qp,q'p')} } ]
\label{eq:CCF_c}
\end{eqnarray}
where the dependence of $\delta_t$ and $\delta_r$ in (\ref{eq:CCF_LOS}) and (\ref{eq:CCF_c}) has been omitted for clarity. The phase difference $\varphi_{L,qp,q'p'}$ of the LOS component is given by
\begin{align}
\label{eq:PD_LOS}
\nonumber
\varphi_{L,qp,q'p'} &=  (p-p')\delta_t \cos(\phi_{L}^t\!-\!\beta_t)\! +\! (q\!-q')\delta_r \cos(\phi_{L}^r\! -\! \beta_r) \\
+&(p\!-\!p')(p'\!+\!p\! - \!M_t-1)\frac{ \delta_t^2sin^2(\phi_{L}^t-\beta_t)}{2D_{TR}}.
\end{align}
The phase difference of the cluster components $\varphi_{c,qp,q'p'}$ is 
\begin{align}
\label{eq:PD_c}
\nonumber
\!\!\!\varphi_{c,qp,q'p'}&= \!(p\!-\!p')\delta_t \cos(\phi_{m_c}^t\!\!\!   -\! \beta_t) \!+\! (q\!-q')\delta_r \cos(\phi_{m_c}^r\!\!\!  -\! \beta_r) \\
+&(p-p')(p'+p - M_t-1) \frac{ \delta_t^2\sin^2(\phi_{m_c}^t-\beta_t)}{2R_c}.
\end{align}
Since the spatial CCFs in (\ref{eq:CCF_LOS}) and (\ref{eq:CCF_c}) depend on the position along the array through the third term in (\ref{eq:PD_LOS}) and (\ref{eq:PD_c}), and they do not only depend on the distance between antenna elements $(p-p')\delta_t$ and $(q-q')\delta_r$, the CIR is non-WSS along the array. Remarkably, the spatial CCF depends on absolute time as can be seen in (\ref{eq:CCF_LOS}). Note that as $R_c$ or $D_{TR}$ become large, the third term in (\ref{eq:PD_LOS}) and (\ref{eq:PD_c}) is reduced to zero and the phase no longer contributes to the non-WSS along the array. 

\section{Simulation results}
\label{sec:Simulations} 
In this section, simulation results of the proposed model will be analyzed. In what follows, delays, cluster powers, AoA, AoD, etc. used in the simulations are generated according to the characteristics of urban macro-cell scenario in \cite{WINNERII-models} unless otherwise specified. The BS and MS are equipped with half-wavelength ($\delta_t=\delta_r=\lambda/2$) equally spaced 128 elements and 10 elements ULAs tilted an angle $\beta_t =\pi/2$ and $\beta_r =\pi/4$ w.r.t. the x-axis, respectively. The frequency of operation is 2.6 GHz. The distance from the BS to the MS $D_{TR}$ is set to 50 m and the distance from the BS to the clusters follows a exponential distribution $R_c \sim exp(\lambda_r)$ with a minimum distance of 20 m and mean $1/\lambda_r = 15$ m.

The parameters controlling the (dis)appearance ($\lambda_{i/v}$) are assigned according to the measurements in \cite{Gao2013}, where clusters with low average power are more likely to disappear. As there is little information based on measurements, all clusters are assigned the same standard deviation $\sigma_c$ regardless of their location. However, since $\mu_c$ determines the cluster average power, it is considered proportional to the cluster average power $\mu_c \propto P_c$ obtained as in (\ref{eq:ClusterPowerWINNNER}). The decorrelation distance of the normal process defining the lognormal power variations over the array has been set to $D_c=0.6$ m. 

\begin{figure}[ht]
\centerline{
\includegraphics[width=0.5\textwidth]{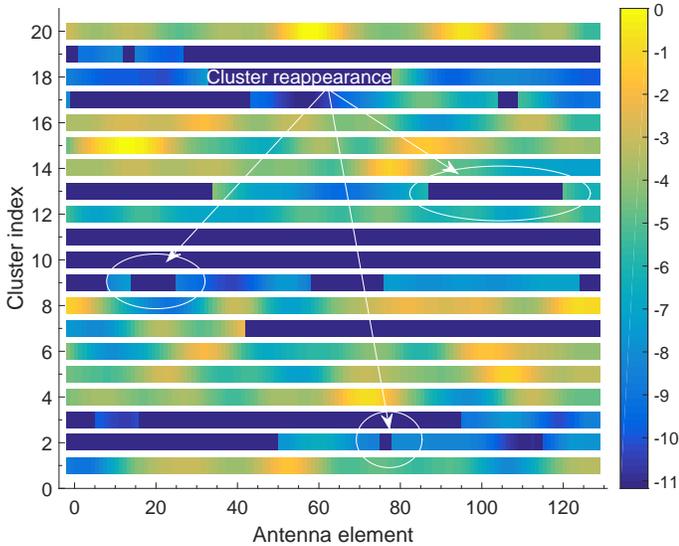} }
\caption{Cluster evolution along the BS array ($\sigma_c=0.2, \mu_c \propto P_c$, $D_c=0.6$ m, $\lambda_{v/i} = 0.01$ m$^{-1}$ for clusters with high average power and $\lambda_{v/i} = 0.5$ m$^{-1}$ for low power ones). }
\label{fig:ClusterEvolution}
\end{figure}

First, an example of cluster evolution along the array is presented in Fig. \ref{fig:ClusterEvolution}. There is a total of 20 clusters of which 16 are visible at the left extreme of the array. Clusters not only appear, disappear and reappear over the array, but their power, represented in  color, smoothly varies along the array according to the lognormal process defined in (\ref{eq:DiscreteLogNormal}). 

\begin{figure}[ht]
\centerline{
\includegraphics[width=0.50\textwidth]{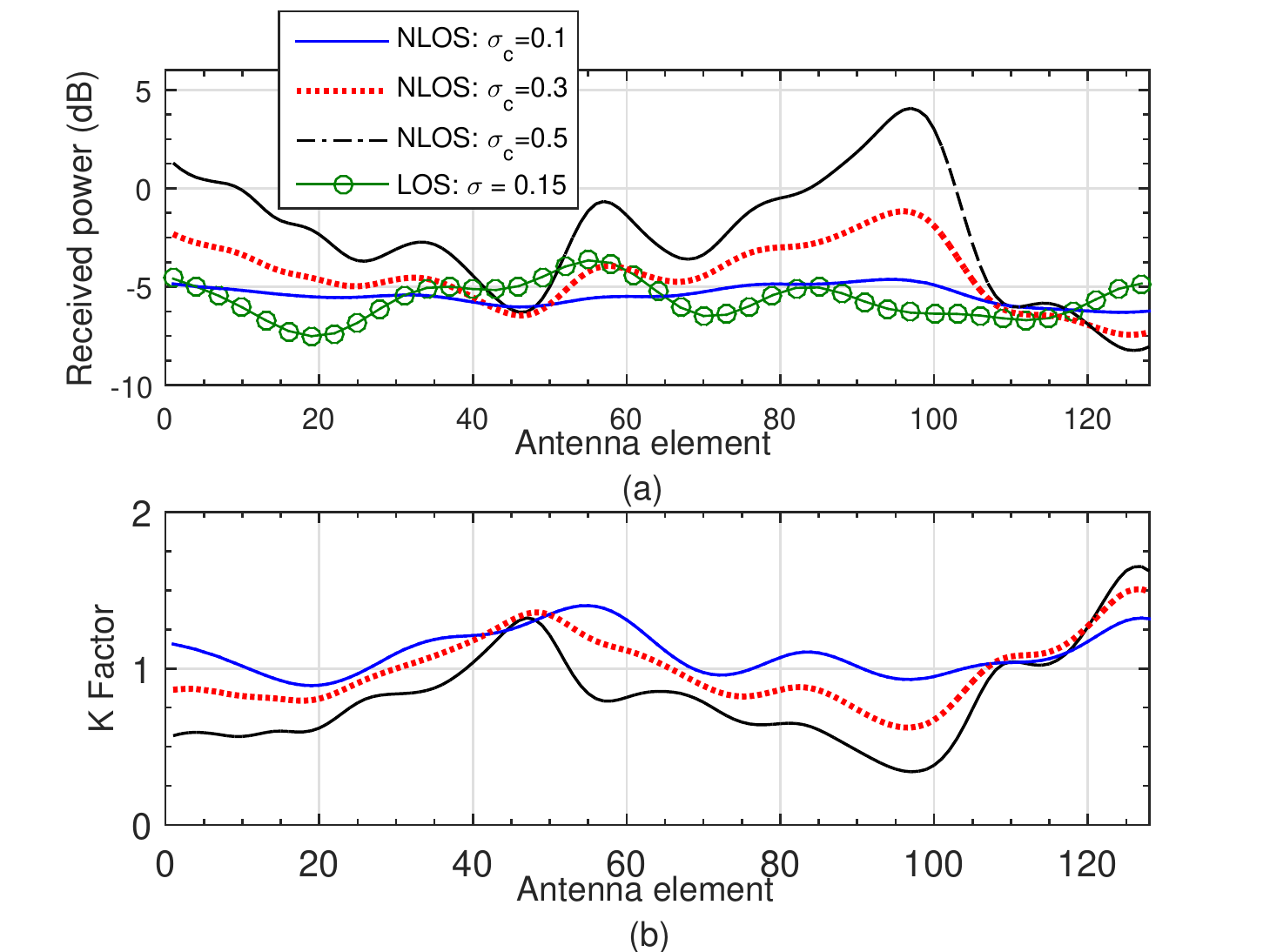} }
\caption{(a) Received cluster power along the array of the LOS/NLOS MPCs for different values of $\sigma_c$. (b) Rician K-factor along the array ($D_c=0.6$ m). }
\label{fig:KFactor}
\end{figure}

As indicated in (\ref{eq:PowerVariations}), the average power of the small scale fading process modeling the channel is variant over large arrays. Consequently, slow variations over the array of the rician K factor (\ref{eq:Kfactor}) are also expected. Fig. \ref{fig:KFactor} (a) shows normalized power variations of the LOS and NLOS MPC for different values of $\sigma_c$ and (b) shows variations of the Rician K factor over the array dimension. As can be easily seen, an increase of the parameter $\sigma_c$ results in higher dynamic range of the average power and K factor. These results show a good agreement with the measurements presented in \cite{Payami2012} and \cite{Gao2013}.

\begin{figure}[ht]
\centerline{
\includegraphics[width=0.46\textwidth]{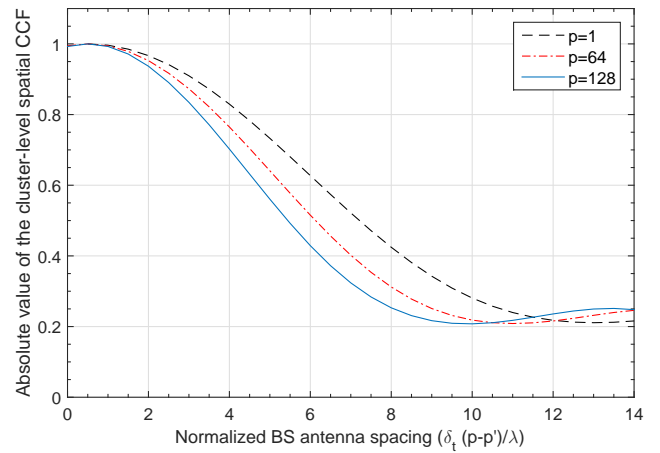} }
\caption{Absolute values of the cluster-level spatial CCFs of the proposed massive MIMO channel model (cluster ASD = $\pi/12$). }
\label{fig:CCF}
\end{figure}

\begin{figure}[htp]
\centerline{
\includegraphics[width=0.5\textwidth]{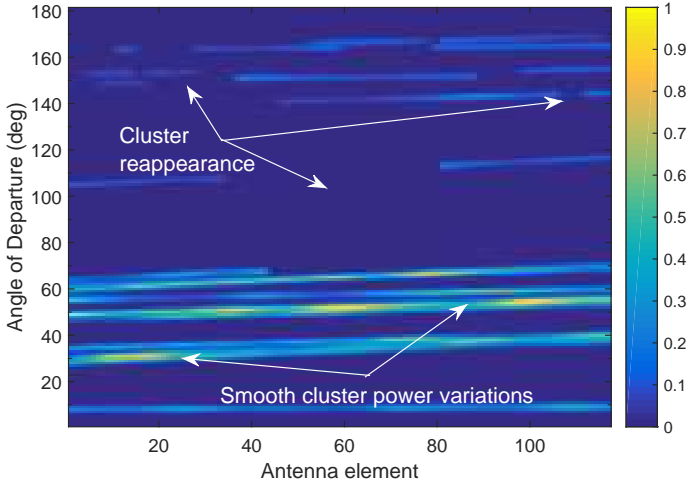} }
\caption{Snapshot of the AoD APS with cluster (re)appearance, angular linear drift and smooth power variations of cluster average power (composite ASD = $\pi/3$).}
\label{fig:APS}
\end{figure}

The absolute value of the cluster-level spatial CCFs of the massive MIMO model derived in (\ref{eq:CCF_c}) is presented in Fig. \ref{fig:CCF} for different antenna elements along the array as a function of the normalized distance between elements.
As it is shown, the spatial CCF does not only depend on the distance between elements but it is different for antennas $p=1$, $p=64$ and $p=128$, i.e., it depends on the absolute position along the array. This demonstrates the non-stationarity of the CIR along the array of the proposed massive MIMO channel model.
Finally, Fig \ref{fig:APS} shows the normalized angle power spectrum (APS) of AoD of the massive MIMO channel model presented. The parameters $\lambda_{i/v}$, $\sigma_c$ and $\mu_c$ are the same as specified in Fig. \ref{fig:ClusterEvolution}. The AoD is estimated using the multiple signal classification algorithm (MUSIC) \cite{Schmidt1986} with a sliding window formed by 12 consecutive antennas shifted one antenna at a time. Only 116 window positions are shown in Fig. \ref{fig:APS}.

It can be seen that the estimated AoDs of MPCs are not constant for consecutive window positions, but different linear variations or drifts of the AoDs over the array are observed as a consequence of the parabolic wavefront described in Section \ref{sec:ParabolicWavefronts}. The slope of the AoDs for every cluster is determined by its distance to the BS. Thus, while the AoD of far clusters is almost constant over the array and its APS is similar to the one observed in conventional MIMO channels, the  drift of the AoDs of closer clusters is more pronounced. Also, low power clusters (dis)appear along the array more often than high power ones as expected. Finally, smooth variations of the average cluster power can be easily seen for high power clusters.

\section{Conclusions}
\label{sec:conclusions}
A novel 2D non-stationary wideband massive MIMO channel model capable of capturing key characteristics of massive MIMO channels has been proposed. It has been shown how parabolic wavefronts are able to model linear drifts of the angles of MPCs on large arrays with lower theoretical and computational complexity. Also, a more flexible and realistic cluster evolution has been considered through the product of two independent Markov two-state and spatial lognormal processes. The main characteristics of massive MIMO channels have been demonstrated such as non-stationarity along the array, cluster (dis)appearance, and smooth evolution of the cluster average power. In future work, we will investigate methods of extracting the parameters proposed in the model from measurements and compare the complexity of the model to that of the existing massive MIMO channel models. 

\section*{Acknowledgment}
The authors gratefully acknowledge the support of this work from the EU H2020 5G Wireless project (Grant No. 641985), the EU FP7 QUICK project (Grant No. PIRSES-GA-2013-612652), the EPSRC TOUCAN project (Grant No. EP/L020009/1), and the Ministry of Science and Technology of China through the 863 Project in 5G (Grant No. 2014AA01A706).
\vspace{-2mm}
\bibliographystyle{IEEEtran}
\bibliography{IEEEabrv,Bibliography}

\begin{thebibliography}{10}
\providecommand{\url}[1]{#1}
\csname url@samestyle\endcsname
\providecommand{\newblock}{\relax}
\providecommand{\bibinfo}[2]{#2}
\providecommand{\BIBentrySTDinterwordspacing}{\spaceskip=0pt\relax}
\providecommand{\BIBentryALTinterwordstretchfactor}{4}
\providecommand{\BIBentryALTinterwordspacing}{\spaceskip=\fontdimen2\font plus
\BIBentryALTinterwordstretchfactor\fontdimen3\font minus
  \fontdimen4\font\relax}
\providecommand{\BIBforeignlanguage}[2]{{%
\expandafter\ifx\csname l@#1\endcsname\relax
\typeout{** WARNING: IEEEtran.bst: No hyphenation pattern has been}%
\typeout{** loaded for the language `#1'. Using the pattern for}%
\typeout{** the default language instead.}%
\else
\language=\csname l@#1\endcsname
\fi
#2}}
\providecommand{\BIBdecl}{\relax}
\BIBdecl

\bibitem{Larsson2013}
E.~G. Larsson, O.~Edfors, F.~Tufvesson, and T.~L. Marzetta, ``{Massive MIMO for
  next generation wireless systems},'' \emph{IEEE Commun. Mag.}, vol.~52,
  no.~2, pp. 186--195, Feb. 2013.

\bibitem{Persson2012}
D.~Persson, B.~K. Lau, and E.~G. Larsson, ``{Scaling up MIMO},'' \emph{IEEE
  Signal Process. Mag.}, vol.~30, no.~1, pp. 40--60, Jan. 2013.

\bibitem{Payami2012}
S.~Payami and F.~Tufvesson, ``{Channel measurements and analysis for very large
  array systems at 2.6 GHz},'' in \emph{Proc. EUCAP'12}, Prague, Czech
  Republic, Mar. 2012, pp. 433--437.

\bibitem{Gao2012}
X.~Gao, F.~Tufvesson, O.~Edfors, and F.~Rusek, ``{Measured propagation
  characteristics for very-large MIMO at 2.6 GHz},'' in \emph{Proc. IEEE
  ASILOMAR'12}, Pacific Grove, CA, USA, Nov. 2012, pp. 295--299.

\bibitem{Gao2013}
X.~Gao, F.~Tufvesson, and O.~Edfors, ``{Massive MIMO channels - Measurements
  and models},'' in \emph{Proc. IEEE ASILOMAR'13}, Pacific Grove, CA, USA, Nov.
  2013, pp. 280--284.

\bibitem{Wu2014}
S.~Wu, C.~X. Wang, E.~H.~M. Aggoune, M.~M. Alwakeel, and Y.~He, ``{A
  non-stationary 3-D wideband twin-cluster model for 5G Massive MIMO
  channels},'' \emph{IEEE J. Sel. Areas Commun.}, vol.~32, no.~6, pp.
  1207--1218, June 2014.

\bibitem{Wu2015a}
S.~Wu, C.-X. Wang, H.~Haas, e.-H.~M. Aggoune, M.~M. Alwakeel, and B.~Ai, ``{A
  non-stationary wideband channel model for massive MIMO communication
  systems},'' \emph{IEEE Trans. Wireless Commun.}, vol.~14, no.~3, pp.
  1434--1446, Mar. 2015.

\bibitem{Liu2012}
L.~Liu, C.~Oestges, J.~Poutanen, K.~Haneda, P.~Vainikainen, F.~Quitin,
  F.~Tufvesson, and P.~Doncker, ``{The COST 2100 MIMO channel model},''
  \emph{IEEE Trans. Wireless Commun.}, vol.~19, no.~6, pp. 92--99, Dec. 2012.

\bibitem{Li2015}
W.~Li, L.~Liu, C.~Tao, Y.~Lu, J.~Xiao, and P.~Liu, ``Channel measurements and
  angle estimation for massive mimo systems in a stadium,'' in \emph{Proc. IEEE
  ICACT'15}, Seoul, South Korea, July 2015, pp. 105--108.

\bibitem{Wang2016b}
C.-X. Wang, S.~Wu, L.~Bai, X.~You, J.~Wang, and C.-L. I, ``{Recent advances and
  future challenges for massive MIMO channel measurements and models},''
  \emph{Sci. China Inf. Sci.}, vol.~59, no.~2, pp. 1--16, Feb. 2016.

\bibitem{WINNERII-models}
K.~Kyosti \emph{et~al.}, ``{WINNER II Channel Models: Part I} channel models,''
  IST-4-027756 WINNER II, Tech. Rep. D1.1.2 V1.2, 2007.

\bibitem{Swind88}
A.~L. Swindlehurst and T.~Kailath, ``Passive direction-of-arrival and range
  estimation for near-field sources,'' in \emph{Proc. IEEE ASSP'88},
  Minneapolis, MN, USA, Aug. 1988, pp. 123--128.

\bibitem{Patzold2012}
M.~Patzold, \emph{{Mobile Radio Channels}}, 2nd~ed.\hskip 1em plus 0.5em minus
  0.4em\relax West Sussex: John Wiley \& Sons, 2012.

\bibitem{Patzold16}
M.~Patzold and C.~A. Gutierrez, ``The wigner distribution of sum-of-cissoids
  and sum-of-chirps processes for the modelling of stationary and
  non-stationary mobile channels,'' in \emph{Proc. IEEE VTC'16 Spring},
  Nanjing, China, May 2016, pp. 1--5.

\bibitem{Papoulis02}
A.~Papoulis, \emph{{Probability, random variables and stochastic processes}},
  4th~ed.\hskip 1em plus 0.5em minus 0.4em\relax USA: McGraw-Hill, 2002.

\bibitem{Schmidt1986}
R.~Schmidt, ``Multiple emitter location and signal parameter estimation,''
  \emph{IEEE Trans. Antennas Propag.}, vol.~34, no.~3, pp. 276--280, Mar. 1986.

\end{thebibliography}

\end{document}